# Better Data Visibility & Data Use Result in Lower Cost and Improved Performance in Medicine Supply Chains

Marasi Mwencha
John Snow, Inc.
Dar es Salaam, Tanzania
marasi_mwencha@jsi.com

James Rosen
Avenir Health
Washington, DC, USA
jrosen@avenirhealth.org

## ABSTRACT
In 2013-2014, Tanzania embarked on a major revamp of the management of its public health supply chains for medicines and other health supplies. These upgrades include the establishment of a national electronic logistics management information system (eLMIS) and the introduction of a Logistics Management Unit (LMU) to use the eLMIS for managing all key public health commodities. This paper describes results from the "round one" evaluation of the impact of those key management upgrades roughly one year after their introduction. The study has three main components: (1) analysis of reporting, data use, management practices, and supply chain outcomes; (2) a cost and cost-effectiveness analysis and (3) a return on investment analysis to measure savings generated by the new systems. The study used a non-experimental pre- and post-design to compare the previous system with the upgraded management system. The quantitative analysis found that stock out rates for all product goods dropped from 32% to 23%, with the frequency of stock-outs greater than 7 days dropping from 24% to 15%. Annual supply chain costs increased from $66million to $76million. Performance improved from the 2014 baseline findings of 68% to 77%, but cost per value of commodities adjusted for performance decreased from 58% at baseline to 50% in year 1.

## 1.INTRODUCTION
In late 2013, Tanzania invested $3.3 million in two key interventions as a major revamp of its public health supply chain management for medicines and other health supplies. The enhancements included introducing a national, electronic logistics management information system (eLMIS) for managing pharmaceutical distribution, and establishing a Logistics Management Unit (LMU) to use the eLMIS data for oversight and performance improvements. Because this investment represents a significant financial commitment to the public health supply chain, a baseline study was undertaken in 2013 and a further analysis in 2015 to evaluate the supply chain performance, cost, and cost-effectiveness of these management upgrades and return on investment on the current system. In addition to providing economic and performance data on supply chain management practices, the analysis carried out also points to areas where these interventions could provide improved supply chain efficiency and generate cost savings.

Development projects must demonstrate real impact and measurable results in order to evaluate the value in financing these initiatives. This study helps answer a fundamental question that ministries of health and donors, alike, are now asking: With limited funding, what management and technology investments can be cost-effective and have a transformative effect on the performance of the supply chain for medicines, vaccines, and other health commodities? Results of this study will provide evidence to inform efforts in other countries seeking to implement similar interventions.

The study had two main components: (1) a baseline and year 1 collection of performance information including supply chain reporting, data use, management practice, and outcomes; and (2) a cost study to provide input for cost-effectiveness analysis. Data collection for the initial analysis of the supply chain took place between August and October 2013 and year 1 data collection took place between April and May 2015. Data came from a range of special surveys applied at different levels of the supply chain, as well as from existing databases. Cost and performance elements were added to existing quarterly surveys and were collected from a nationally representative sample of 220 health facilities. The evaluation analyzed results related to three major logistics systems, which handle the vast majority of products in Tanzania: the Integrated Logistics System (ILS), through which essential medicines are managed; the antiretroviral drugs (ARV) logistics system; and the expanded program on immunization (EPI), which manages vaccines. ILS data was segmented to allow analysis of Family Planning and Malaria product groups.

## 2.DEVELOPMENT IMPACT
As in many other developing countries, the management of the public health supply chain in Tanzania evolved along product lines associated with specific public health programs. The result was a fragmented, uncoordinated management system with separate structures for HIV, TB, vaccines, essential drugs, and contraceptives. A key management function—information management—evolved along product groupings. These separate, and largely paper-based logistics management information systems (LMIS), did not generate accurate, quality, and timely data. The lack of data visibility contributed to poor system performance that raised costs, hampered stock availability, and made it harder to serve the millions of Tanzanians who rely on public health facilities.

To address these problems, the Ministry of Health Community Development Gender Elderly and Children (MOHCDGEC) set up the LMU, a management structure responsible for coordinating, monitoring, and supporting all of the logistics activities for various commodity groups. LMU staff identify supply chain bottle-necks, develop solutions for those challenges, and implement corresponding interventions. Following the implementation of the LMU, the MOHCDGEC facilitated the rollout of a nationwide eLMIS to consolidate existing electronic and paper-based systems; such consolidation is a challenge faced by many developing nations. The eLMIS is expected to improve data visibility by providing accurate, quality, and timely data. The LMU will be responsible for actively managing, analyzing and using data generated from the eLMIS.

The MOHCDGEC and other stakeholders also wanted to understand the cost-effectiveness of the current system and what portions of the system might see cost savings and performance improvements after the LMU and eLMIS interventions.

## 2.1. RESULTS

From the quantitative analysis that was conducted, stock out rates—defined as zero availability of a given product at the point of service—on average across all product groups dropped from 32% to 23%, with the frequency of stock-outs greater than 7 days dropping from 24% to 15%. Appropriate inventory levels at health facilities increased slightly from 18% to 20%, indicating that increases in stock availability do not appear to be a result of overstocking.

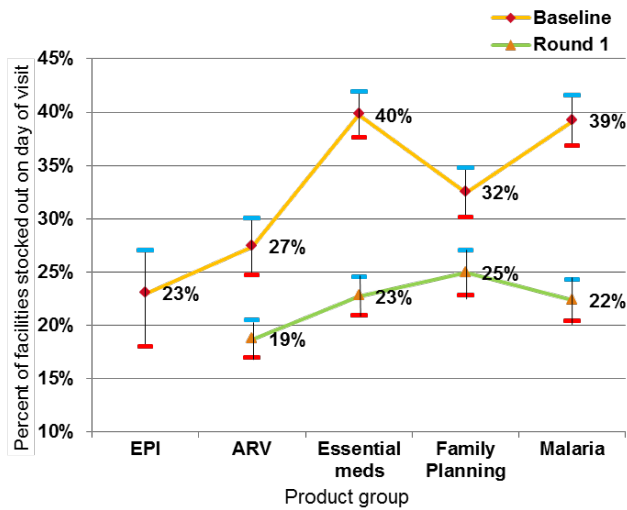

**Figure 1: Facilities stocked out on day of visit**

The levels of expiries at the facility also improved, decreasing by 0.1-0.4%. With respect to cost data, the annual cost of the supply chain increased from $66million to $76million. The throughput of the supply chain also increased from $168million to $197million. Despite this increase in costs and throughput, performance of the supply chain improved from the 2014 baseline findings of 68% to 77%. The cost per value of commodities adjusted for performance shows a decrease from 58% at baseline to 50% in year 1. A time-series analysis drawing on three years of quarterly data also found statistically significant positive changes in these performance indicators.

## INTERPRETATION AND RECOMMENDATION

Findings from the year 1 data collection indicate that the information and management upgrades had a positive impact on key supply outcomes, especially stock availability. Both the incidence and the duration of stock-outs decreased, while inventory levels remained relatively constant. While the upgraded supply chain system costs were higher, initial results indicate that it is operating more efficiently; the implementation of the LMU and eLMIS reduced overall expiry rates, resulting in savings due to lower wastage. Moreover, these management upgrades also appear to have generated significant savings to the government despite the supply chain challenges that the country experienced at the same time.

It is however important to note that while performance and cost data are generally supportive and justify the level of investment to improve the supply chain, many of the gains noted may be fragile. For example, there was little movement in some management practice indicators between baseline and round 1 data collection. As such, it will be important to conduct further data collection surveys to discern the true impact and performance, cost and cost-benefit of these interventions.

## 3. CONCLUSION

The improved visibility of supply chain data provided by the eLMIS was a critical step to improving performance and reducing costs. However, technology-driven data visibility alone is insufficient. For real change to happen, the data must be analyzed and used for routine and strategic decisions and for continuous quality improvement. The LMU established the core team of dedicated data users who have the mandate to analyze supply chain data and affect change.